\documentstyle[12pt]{article}
\begin{document}
\date{}
\title{Phase-Boundaries near Critical End points: Applications to
 Cross-linked Copolymers}
\author{Edilson Vargas and Marcia C. Barbosa \\
Instituto de F\'{\i}sica, Universidade Federal
do Rio Grande do Sul\\ Caixa Postal 15051, CEP 91501-970
, Porto Alegre, RS, Brazil}
\maketitle
\begin{abstract}
The phase behavior of a cross-linked  polymer blend 
made of two incompatible species, $A$ and $B$,  of different chemical
nature is analyzed. Besides a  homogeneous
phase, this system also exhibits two  microphases and a 
phase of total segregation. The transition between
the homogeneous and the microphase is continuous along
a $\lambda$-line;  a first-order
phase boundaries separate the microphase and the disordered
phase from the phase of complete segregation.
The critical line meets the first-order phase boundaries at
an end point. 
Scaling arguments indicate that, close to {\it any} end point, the equations for the
first-order phase boundaries exhibit  nonanaliticities associated
with the singularities present at the thermodynamic functions
near  the critical line.
Explicit expressions for the phase boundaries near the end point  for 
a cross-linked polymer mixture are obtained and checked for singularities. 
\end{abstract}
\section{\bf Introduction}

Polymer mixtures were usually phase separated \cite{De77}. When the
chains are long, the translational entropy of the chain is small and
any chemical difference between the two species leads to a repulsion
and consequent phase separation  at a temperature $T=T_0$. One way 
to prevent phase-separation is to cross-link the system at high temperatures 
and then brought it to the coexistence region. A competition between the 
natural tendency of the system to phase separate and the elastic forces
due to the presence of the cross-links is established. As a result of
this competition, there is a microphase separation \cite{De79}.
For strongly cross-linked system, the distance between two cross-links
is fixed and  the cross-links are permanent. In this case, the 
microphase separation happens at a temperature  lower than
the temperature for which complete segregation would had occurs $T_m<T_0$ \cite{De79}.

When the number of cross-links is not too large, the
distance between cross-links it is not
fixed and the position of each $A$-$B$ link
fluctuates in space. Then, as these inhomogeneities in the links are to be taken
taken into account, the phase-diagram of the system changes quite drastically
\cite{Sc94}\cite{St94}\cite{Va97}. Besides the homogeneous phase and
the microphase, one also finds a phase of complete segregation.
This phase  
is separated from the homogeneous and microphase by  first-order phase boundaries.
The critical line between the disordered and the
microphase meets these phase boundaries at an end point where
the three phases coexist.

It is well known that in the vicinity of continuous transitions, singularities
in the thermodynamic functions such as compressibility, specific heat
and density should be expected. These nonanalyticities are usually expressed through 
universal critical exponents and universal amplitude ratios that one can
measure. However, some time ago \cite{Ba1}\cite{Ba2}, it was shown, through 
a phenomenological argument, that in the vicinity of an end point, singularities
in the first order phase boundaries should also be expected
and that these singularities could also be expressed in terms of 
universal exponents and universal amplitude ratios. Nevertheless, this
assumption  is based on scaling arguments  and  needs to be checked. This
was already done in the spherical model \cite{Ba3}. Unfortunately
the spherical model has no physical realization and, consequently, 
we indeed need to check our theory in an more realistic model. 

In this sense, our aim in this paper is to show that these singularities
are actually present in the first-order phase boundaries of
the polymer mixture discribed above. In order to do so, on
basis of an effective mean-field theory
that takes into account anisotropy between the
two species and inhomogeneities in the cross-links,
 we calculate the phase boundaries near
the end point. The remainder of this paper is organized as
follows. In sec. $2$, we briefly review the scaling arguments 
that led us to the conclusion that near the end points singularities
in the phase  boundaries might exist. Accepting the plausibility but
the uncertainty of this conclusion, in sec. $3$, we test
them in a  model for polymer mixture. Our results are summarized in sec. $4$.

\section{\bf Phase Boundaries near Critical End points}

Before analyzing  the phase-diagram of the copolymers, it would be
helpful to  review the theory for singularities
in the phase boundaries near end points
\cite{Ba1}-\cite{Ba4}. 
In the space $(T,g,h)$ ( here $g$ is a intensive
parameter that can be a pressure, external field, chemical
potential, etc), an end point is the
locus
where two phases, $II^+$ and $II^-$, become critical
in the presence of a third noncritical 
phase, $III$ (see Fig.$1$).
Now, what is the form of the
phase boundary $g_{\rho}(T,h)$ that separates the critical phases from
the spectator phase
when $(T,h)$ pass through $(T_e,h=0)$?
At the end point $g$ has a definite value, $g_{e}$, and
so do the slopes $g_{1}\equiv dg_{\rho}/dT\mid_e$ and
$g_{2}\equiv dg_{\rho}/dh\mid_e$. 
Thus, it seems likely that most people would assume that
$g_{\rho}=g_{e}+g_{1}t+g_{2}h+...$ 
and further {\it  analytic terms} in the Taylor
series expansion in powers of 
of $ t \equiv (T-T_{e})/T_{e})$ and $h$.
However, this is wrong. Fisher and Barbosa  {\it (FB)} \cite{Ba1}
have shown that the full expression for $g_{\rho}$ 
given by
\begin{eqnarray}
\label{pb1}
g_{\rho}&=& g_{e}+g_{1} t + g_{2} h - X_{\pm}\mid t \mid^{2-\alpha}
-Y\mid t \mid^{\beta}\mid h \mid - Z_{\pm} \mid t \mid^{-\gamma} h^2+ ...
\end{eqnarray}
contains nonanaliticities
related with the singularities at the
critical line. Here  $+$ means the disordered phase ($I$) and the  $-$
refers to the ordered phases ($II^{\pm}$), the exponents 
$\alpha,\beta$ and  $\gamma$ are 
the specific heat,
the order parameter and the susceptibility critical 
exponents associated with   $\lambda$ line.
Similarly the amplitudes $X_{\pm}$, $Y$ and $Z_{\pm}$
are universally related to the critical amplitudes by
\begin{eqnarray}
\label{r}
\frac{X_{+}}{X_{-}}&=&\frac{A_{+}}{A_{-}}\\ \nonumber
\frac{Z_{+}}{Z_{-}}&=&\frac{C_{+}}{C_{-}}\\  \nonumber
\frac{X_{+}Z_{+}}{Y^2}&=&\frac{A_{+}C_{+}}{(2-\alpha)(1-\alpha)B^2}
\end{eqnarray}
where $A_{\pm}$, $B$ and $C_{\pm}$ are the critical amplitudes
for  the specific heat, for the order parameter and  for the susceptibility
near the critical line respectively. 

But this is not the only nonanaliticities present in this
phase boundary. Indeed the intersection of the surface $\rho$
with the plane $t=0$  is a curve that near the end point
can be written as 
\begin{eqnarray}
\label{pb2}
g_{c}(h)&=&g_{e}-Y_{c}\mid h \mid^{1+1/\delta}
\end{eqnarray}
where $\delta=\beta\Delta$ and where $\Delta$ is the gap exponent
of the critical region.
Here, as before, the coefficient $Y_{c}$ is also related to the
critical amplitudes by
\begin{eqnarray}
\label{ryc}
\frac{Y^{\delta-1}Z_{+}}{Y^{\delta}_{c}}&=&(\frac{\delta+1}{\delta})^{\delta}
\frac{B^{\delta-1}A_{+}}{B^{\delta}_{c}}      .
\end{eqnarray}

Therefore, if the $\lambda$-line
would be classical, 
{\it FB's} theory predicts that the phase boundary $g_{\rho}$ 
given by Eq.~(\ref{pb1})
would have two nonanalytic terms with exponents $\beta=1/2$ and
$\gamma=1$, while the phase boundary $g_{c}$ 
given by Eq.~(\ref{pb2}) would have
one singular term with exponent $\delta=3$. These 
are the usual mean-field critical exponents. Besides, {\it FB} 
approach also predicts
that the 
coefficients $Z_{\pm}$, $Y$ and $Y_{c}$ would be universally related
to the critical amplitudes $C_{\pm}$, $B$ and $B_{c}$ by
\begin{eqnarray}
\label{rmf}
\frac{Z_{+}}{Z_{-}}&=&\frac{C_{+}}{C_{-}}=2 \\  
\frac{Y^{\delta-1}Z_{+}}{Y^{\delta}_{c}}&=&(\frac{\delta+1}{\delta})^{\delta}
\frac{B^{\delta-1}A_{+}}{B^{\delta}_{c}}=(\frac{4}{3})^{3}.     \nonumber
\end{eqnarray}

\section{\bf Phase-Diagram of Crossliked Copolymers }

In this section, we investigate the melt of a non compatible and cross-linked mixture
of  polymers $A$ and $B$ using the Landau-Ginzburg-Wilson-de Gennes'
Hamiltonian \cite{De77}\cite{Va97}
\begin{eqnarray}
\label{H}
\beta H&=&\int_{}^{} d^{3}r 
\{ \frac{(a\nabla \phi(r))^2}{48}+\frac{\tau}{2}\phi(r)^2+
u\phi(r)^4-h(r)\phi(r)+\frac{C(r) P(r)^2}{2}\}
\end{eqnarray}
where $a$ is the size of one monomer and where the order parameter $\phi(r)$ 
is given in terms of the  local fluctuations of the density of
each specie, $\phi_A(r)$ and $\phi_B(r)$ by
\begin{eqnarray}
\label{phi}
\phi_A(r)&=&\frac{1}{2}(1+l(r)+\phi(r))\nonumber \\ 
\phi_B(r)&=&\frac{1}{2}(1-l(r)-\phi(r)) .
\end{eqnarray}
Here $\langle \phi_A(r) \rangle=1/2(1+\langle l(r) \rangle)$ 
and $\langle \phi_B(r)\rangle=1/2(1-\langle l(r) \rangle)$ are the
volume fractions of each  type of polymer. A nonzero value of
$\langle l(r) \rangle$ allows for different volume
fractions of each specie. The  term linear in $\phi (r)$ in Eq.~(\ref{Heff})
contains the difference in the chemical potential
between the two types of polymer.
In each cross-link, two  
monomers one belonging to the specie $A$ and 
another to the specie $B$ are tied together. However they can be
displaced slightly, leading to an elastic
``polarization'' given by
\begin{eqnarray}
\label{P}
\vec{P}&=&\frac{1}{V}(\sum_{i\in A}^{}\vec{r}_i - \sum_{j\in B}^{}\vec{r}_j)
\end{eqnarray}
 where $\vec{r}_i$ is the position of the $i$ monomer at a polymer
 of type $A$  while $\vec{r}_j$ is the position of the $j$ monomer
 of type $B$ and where $V$ is the total volume of the system.
 In the same way
 that for  the electrostatic case, polarization and charge are not
 independent quantities
 , here the elasticity and the volume fraction of each specie
 are also related by $\nabla \cdot \vec{P}=\phi(r)+l(r)$.

Consequently, the last term at the Hamiltonian Eq.~(\ref{H})
contains the elastic contribution  associated with 
the cross-links. For simplicity, we assume that
this term has a quadratic form
that resembles the energy of a spring system. 
 $C(r)$ is the internal rigidity is given by
given by
$C(r)=C_0\sum_{\vec{r}_i}^{}\delta(\vec{r}-\vec{r}_i)$ where
here $\{\vec{r}_i\}$ correspond to coordinates of $N_c$ cross-links
distributed in the volume $V$
 according a Poisson distribution
characterized by
$\langle C(r_1)C(r_2) \rangle = C_{0} \langle C(r_{1}) \rangle\delta(r_1-r_2)$.

Since the cross-links are not permanent, they can open and close. Consequently
 the disorder is
assumed to be annealed. Then, the resulting effective hamiltonian
obtained after averaging over this distribution is given by
\begin{eqnarray}
\label{Heff}
\beta H_{eff}&=&\int_{}^{} d^{3}r 
\{ a^2\frac{(\nabla \phi(r))^2}{48}+\frac{\tau}{2}\phi(r)^2+
u\phi(r)^4-h(r)\phi(r) \nonumber \\
&+&\frac{1}{n}\int_{}^{} d^3 r [1- e^{C_{0} P^{2}/2}]\}.
\end{eqnarray}
where  $n=V/N_c$ ( actually
$1/n$ is the density of cross-links). If the gel is very dense, $n$ will 
be related to the average number of monomers between two cross-links.

Now, one has to eliminate 
$\vec{P}$ in favor of $\phi)(r)$,
 using  the constraint $\nabla \cdot \vec{P}=\phi(r)+l(r)$. 
Then, the  expression for the free energy $\beta F_{eff}=-\ln e^{-\beta H_{eff}}$
 can be evaluated at the
mean-field level by taking the saddle point approximation,
what leads to 
\begin{eqnarray}
\label{F}
\beta F_{eff}&=&\frac{1}{2}[\tau+\frac{(q_ca)^2}{24}]\psi_{q_c}\psi_{-q_c}
+u\psi_{q_c}^2\psi_{-q_c}^2
-h_{-q_c}\psi_{q_c}+\frac{1}{n}[1-e^{- cn\psi_{q_c}\psi_{-q_c}/(2q_c^2)}]\nonumber
\\
\end{eqnarray}
where the expressions for
 $\psi_{q_c}$ and $q_c$ are given by the equations
\begin{eqnarray}
\label{dF1}
\frac{\partial \beta H_{eff}}{\partial{\phi(q)}}\mid_{\phi_q=\psi_{q_c},q=q_c}
&=&[\tau+\frac{(q_c a)^2}{24}]\psi_{-q_c}+4u\psi_{-q_c}^2\psi_{q_c}
-h_{-q_c}\nonumber\\
&+&\frac{c}{q_c^2}\psi_{-q_c}e^{- cn\psi_{q_c}\psi_{-q_c}/(2q_c^2)}
\end{eqnarray}
and
\begin{eqnarray}
\label{dF2}
\frac{\partial \beta H_{eff}}{\partial{q}}\mid_{\phi_q=\psi_{q_c},q=q_c}&=&
a^2\frac{q_c }{24}\psi_{q_c}\psi_{-q_c}-\frac{c}{q_c^3}
e^{-cn\psi_{q_c}\psi_{-q_c}/(2q_c^2)}
\end{eqnarray}
and where $c=C_0/n$.
From the above equations we can see that the system exhibits four possible
phases :

$(a)$ \underline{phase $I$}, a {\it homogeneous phase} 
where $\psi_I\rightarrow 0$ as 
$h_{q_{I}}\rightarrow 0$ and  where $q_{c}=q_I\neq 0$;;

$(b)$\underline{ phases $II_+$ and $II_-$}, two
microphases where
{\it partial segregation} occurs, where $\psi_{II}\not\rightarrow 0$ as 
$h_{q_{II}}\rightarrow 0$ and where $q_{c}=q_{II}\neq 0$.

$(c)$ \underline{phase $III$},
a {\it complete segregated phase},  where $\psi_{III}\not\rightarrow 0$ as 
$h_{q_c}\rightarrow 0$ and where $q_c=q_{III}=0$;

The free energy associated with each one of these phases is given by
\begin{eqnarray}
\label{I}
\beta F_I&=&\frac{1}{2}[\tau+\frac{(a q_I)^2}{24}]
\psi_I^2+u\psi_I^4+\frac{1}{n}[1-e^{-[cn\psi_I^2/(2q_I)]}]
-h_{-q_I} \psi_I
\end{eqnarray}
for the phase $I$,
\begin{eqnarray}
\label{II}
\beta F_{II}&=&\frac{1}{2}[\tau+\frac{(aq_{II})^2}{24}]\psi_{II}^2
+u\psi_{II}^4+\frac{1}{n}[1-e^{-[cn\psi_{II}^2/(2q_{II})]}]-h_{-q_{II}} \psi_{II}
\nonumber \\ 
\end{eqnarray}
for phase $II$ and
\begin{eqnarray}
\label{III}
\beta F_{III}&=&\frac{\tau}{2}\psi_{III}^2+\frac{u}{4}\psi_{III}^4+\frac{1}{n}
\end{eqnarray}
for the phase $III$. Here 
the values of $\psi_I,\psi_{II},\psi_{III},q_{I}$ and $q_{II}$
are given by the saddle point solutions of Eq.~(\ref{dF1}) and 
Eq.~(\ref{dF2}).

By comparing the above free energies we find the phase-diagram 
illustrated
in fig. $2$ that goes as follows \cite{Va97}.  At high values of $\tau$ that
here plays the hole of  temperature, the two species
are mixed at phase $I$. For  strongly cross-linked system,
low values of $n$,   as
the temperature is decreased, one finds at $\tau=\tau_e$ 
a continuous phase transition  
to  the microphase ( phase $II$). If the temperature 
is decreased even further,
one meets a first-order phase boundary  between the
microphase, phase $II$, and a state where
the two species are completely segregated, phase $III$. When
 the system is
weakly cross-linked, the number of monomers between two cross-links, $n$,
is large and  consequently,  as the temperature
is decreased,  one finds a first-order phase transition between the
homogeneous state, phase $I$, and 
the state where the segregation is total, phase $III$.  The three phases, $I$,
$II$ and $III$ meet at 
the end point $e$ at $(\tau=\tau_e,n=n_e,h=0)$.

Then, in order to check if our predictions summarized at the 
Eq.~(\ref{pb1}), Eq.~(\ref{pb2}), Eq.~(\ref{r}) and Eq.~(\ref{ryc}),
 are correct, we will
obtain the expression for the phase boundary $\rho$ near the end point.
First, by equating the free energy of the homogeneous phase given by
Eq.~(\ref{I}) to the free energy of the phase $III$ given by Eq.~(\ref{III})
we find the expression 
\begin{eqnarray}
\label{rho+}
\tau_{\rho +}&=& \tau_{e}+g_{1} (n-n_e)  - X_{+}\mid n-n_e \mid^{2-\alpha}
- \frac{1}{2}Z_{+} \mid n-n_e \mid^{-\gamma} h^2
\end{eqnarray}
for  the phase boundary  $\rho_{+}$ near the end point
when $n>n_e$.
Here  the exponents  are $\alpha=0$ and $\gamma=1$ while
the non-universal  coefficients are  
\begin{eqnarray}
\label{c+}
g_{1}&=& \frac{c^{3/2}}{u}    \\ \nonumber
X_{+}&=& \frac{c^{5/2}}{4u^2}        \\ \nonumber
Z_{+}&=& \frac{u^2}{c^2}          .
\end{eqnarray}

 Similarly, equating the free energy of the microphase,
phase $II$, given by Eq.~(\ref{II}) to the free energy of the phase $III$
given by Eq.~(\ref{III}), 
we obtain the expression for the phase boundary $\rho_{-}$ 
near the  end point when $n<n_e$
namely
\begin{eqnarray}
\label{rho-}
\tau_{\rho -}&=& \tau_{e}+g_{1} (n-n_e)  - X_{-}\mid n-n_e \mid^{2-\alpha}
-Y\mid n-n_e \mid^{\beta}\mid h \mid \nonumber \\
&- &Z_{-} \mid n-n_e \mid^{-\gamma} h^2  
\end{eqnarray}
where $\alpha=0$, $\beta=1/2$ and $\gamma=1$,
where the coefficient $g_{1}$ is the same as the one
obtained in Eq.~(\ref{c+}) 
as we have predicted ( see Eq.(\ref{pb1})) and where
\begin{eqnarray}
\label{c-}
X_{-}&=&\frac{5c^{5/2}}{4u^2}          \\ \nonumber
Y&=& \sqrt{2}c^{1/4}             \\ \nonumber
Z_{-}&=&\frac{u^2}{2c^2}          .
\end{eqnarray}

Now, one can easily verify that the ratios  Eq.~(\ref{r})
built on 
basis of  Eq.~(\ref{c+}) and Eq.~(\ref{c-}) are universal and given by
the values predicted at
Eq.~(\ref{rmf}).

The intersection of the plane $n=n_e$ with the surface $\rho$ is
a curve given by
\begin{eqnarray}
\label{rhoc}
\tau_{c}(h)&=&\tau_{e}-Y_{c}\mid h \mid^{1+1/\delta}
\end{eqnarray}
where $\delta=3$ as we have predicted and where
\begin{eqnarray}
\label{cc}
Y_{c}&=& \frac{4 (2u^{2})^{1/3}}{3 c^{1/2}}         .
\end{eqnarray}
Using the value above for the coefficient
$Y_c$ together with Eq.~(\ref{c+}) and Eq.~(\ref{c-})
we obtain the universal values predicted by Eq.~(\ref{rmf}).

\section{\bf Results and Conclusions}

Scaling arguments indicate that near any critical end  point the phase boundaries
must exhibit a nonanalyticities related to the singularities of the critical
line.  
 
In order to verify if this prediction based in a phenomenological approach
is actually correct, we have looked for a realistic system that exhibits an end point.
Therefore,
we have studied the phase behavior of a mixture of two 
incompatible polymer
species, $A$ and $B$, that at high temperatures are cross-linked.
The phase-diagram for an asymmetric polymer blend with
 inhomogeneities in the cross-links, illustrated
in figure $2$, exhibits a critical end point where the
four phases present in this system, a homogeneous phase, two microphases 
and a phase of complete segregation,  meet. 
Then, using a mean-field theory,  
expressions for the phase boundaries  separating the critical phases from 
the non-critical phase were derived. Finally we checked that these
equations exhibit nonanalyticities associated with the
mean-field critical
 exponents of the specific heat, $\alpha=0$, of the order parameter
 , $\beta=1/2$, of the isothermal compressibility, $\gamma=1$ and 
 of the order parameter at the critical temperature $\delta=3$.  
We also verified that the coefficients $X_{\pm}, Z_{\pm}, Y$ and
$Y_c$  are universally related to the
critical amplitudes $A_{\pm}, C_{\pm}, B$ and $B_{c}$ as was
predicted  in Eq.~(\ref{r}) and
in Eq.(\ref{ryc}). The universal ratios assume the classical
values given by Eq.~(\ref{rmf}). 
\bigskip
\bigskip

\centerline{\it ACKNOWLEDGMENTS}

\bigskip

This work was supported in part by CNPq - Conselho Nacional de
Desenvolvimento Cient\'{\i}fico e Tecnol\'ogico and FINEP -
Financiadora de Estudos e Projetos, Brazil.

\bigskip
\bigskip

\centerline{\bf FIGURE CAPTION}

\bigskip
\bigskip

\noindent Figure $1$.  Schematic phase-diagram $t \times g \times h$.  The phase
$I$ is the disordered phase, phase $II_{\pm}$ are the two ordered phases 
and phase $III$ is the noncritical phase. The  line $\lambda$, dashed line, is 
a continuous transition,  the planes $\rho$ 
and $\eta$ are  first-order phase boundaries and $e$ locates the end point.
The first-order lines $\sigma_{\pm}$ are
the intersection of the surface $\rho$ with the plane $h=0$.  

\bigskip
\bigskip
\noindent Figure $2$ .  Phase-diagram $n \times \tau \times h$ for 
a $A$-$B$ polymer bend. The phase
$I$ is the homogeneous phase, phases $II_{\pm}$ are the microphases
and phase $III$ is the phase where the segregation is total. The  line $\lambda$, dashed line, is 
a continuous transition,  the planes $\rho$ 
and $\eta$ are  first-order phase boundaries and $e$ locates the end point.
The first-order lines $\sigma_{\pm}$ are
the intersection of the surface $\rho$ with the plane $h=0$.

\end{document}